\documentclass{moriond}
\usepackage[utf8]{inputenc} 
\def\be{\begin{equation}}
\def\ee{\end{equation}}
\def\bea{\begin{eqnarray}}
\def\eea{\end{eqnarray}}

\newcommand{\gev}{\mathrm{\,GeV}}

\newcommand{\smsq}{{m_{\tilde{q}}}}

\newcommand{\smgo}{{m_{\tilde{g}}}}
\newcommand{\sq}{{\widetilde{q}}}

\renewcommand{\L}{{\mathrm{L}}}
\newcommand{\R}{{\mathrm{R}}}

\newcommand{\Tinv}{\mathrm{T2}_\infty}

\newcommand{\tmgo}{\mathrm{T2}_\smgo}
\newcommand{\s}[1]{\tilde{#1}}

\newcommand{\mne}{m_{\s\chi^0_1}}

\newcommand{\Photo}{}

\begin{document}
\vspace*{4cm}
\title{LIMITS AND FITS FROM SIMPLIFIED MODELS~\footnote{Talk presented at the 50th Rencontres de Moriond (EW session), La Thuile, March 18th 2015.}}

\author{ JORY SONNEVELD }
%\author{Lisa Edelh\"{a}user, Jan Heisig, Michael Kr\"{a}mer, Lennart Oymanns, Jory Sonneveld, Wolfgang Waltenberger }
\address{Institute for Theoretical Particle Physics and Cosmology, RWTH Aachen, Sommerfeldstra\ss e 16, \\
  52074 Aachen, Germany}
\maketitle
\abstracts{
  An important tool for interpreting LHC searches for 
new physics are simplified models. They are characterized by a 
small number of parameters and thus often rely on a 
simplified description of particle production and decay dynamics.
We compare the interpretation of current LHC searches for hadronic jets plus missing energy signatures 
within simplified models with the interpretation within complete supersymmetric and same-spin models of quark partners.
We found that the differences between the mass limits derived from a simplified model and 
from the complete models are moderate given the current LHC sensitivity. 
We conclude that simplified models provide a 
reliable tool to interpret the current hadronic jets plus missing energy searches at the LHC in a 
more model-independent way.}

\section{Introduction}
In order to cover a broad part of BSM theories' parameter space, current searches for new physics at the LHC use simplified models motivated by supersymmetry to quantify their results~\cite{ArkaniHamed:2007fw,Alwall:2008ag,Alves:2011wf,Alves:2011sq,Chatrchyan:2013sza}. Several recently developed tools~\cite{Kraml:2013mwa,Drees:2013wra,Papucci:2014rja,Kraml:2014sna} use simplified model results to enable one to test BSM theories against LHC results. It is thereby assumed that upper limits calculated from signal efficiencies for simplified models are mostly unchanged compared to more realistic, more complicated models that include more particles or even have particles with a different spin.

One simplified SUSY model for squarks called \texttt{T2}, shown in figure \ref{fig:t2}, includes light-flavor squarks, where squarks decay to a quark and a lightest supersymmetric stable particle (LSP). Gluinos (as all other supersymmetric particles except the LSP) are decoupled, and it is assumed that the quark partners produced in this simplified model are scalar particles.

We investigated the effects of adding a finite-mass gluino to the simplified model of squarks used by the experimental collaborations, as well as the influence on mass limits upon changing the spin assumption.

\section{Production of squarks at the LHC}
\begin{figure}[]
\begin{minipage}[t]{0.45\textwidth}
%\begin{figure}[]
  \begin{center}
    \includegraphics[width=0.5\textwidth]{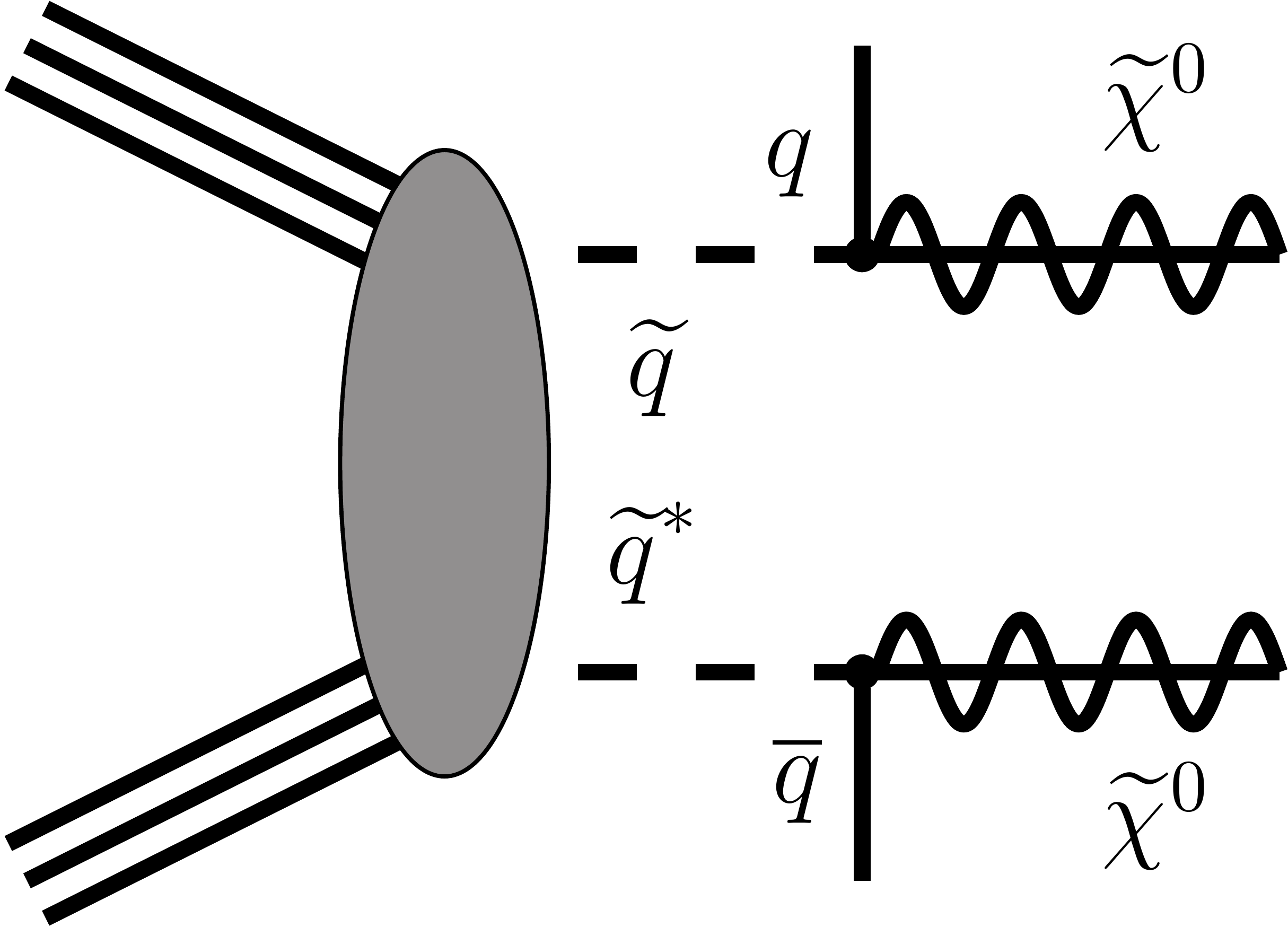}
  \end{center}
  \caption{The simplified model \texttt{T2} for squark production at the LHC as used by the experimental collaborations. The gluino is decoupled.}
  \label{fig:t2}
%\end{figure}
\end{minipage}
\hspace{0.5cm}
%\begin{minipage}[t]{0.1\textwidth}
%\color{white}{notext}
%\end{minipage}
\begin{minipage}[t]{0.45\textwidth}
%\begin{figure}[]
  \begin{center}
    \includegraphics[width=0.45\textwidth]{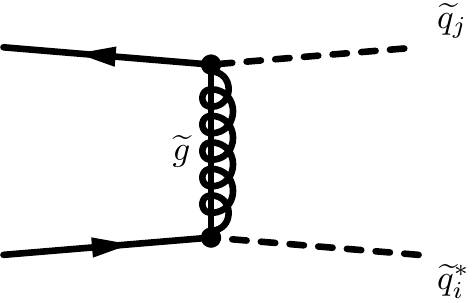}
    \includegraphics[width=0.45\textwidth]{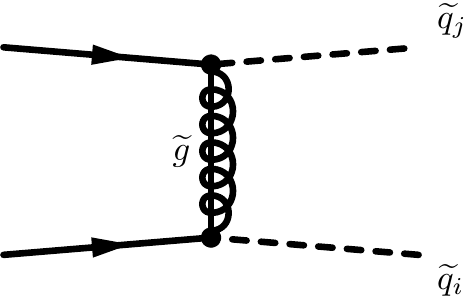}
  \end{center}
  \caption{The $t$-channel gluino diagrams that contribute only when finite-mass gluinos are present.
  }
  \label{fig:tchannel_gluino}
%\end{figure}
\end{minipage}
\end{figure}

\begin{figure}[]
  \begin{center}
\includegraphics[width=0.15\textwidth]{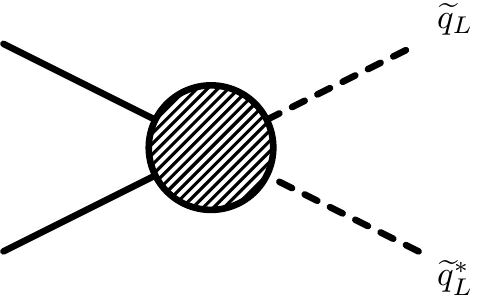}\hspace{2ex}\raisebox{3ex}{=\hspace{2ex}$\Bigg\{$}
\includegraphics[width=0.15\textwidth]{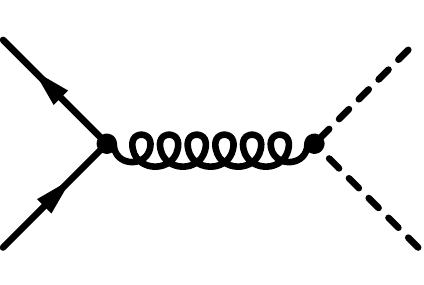}
\includegraphics[width=0.15\textwidth]{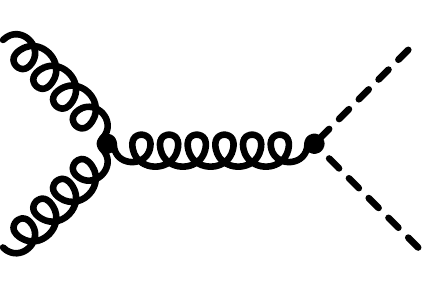}
\includegraphics[width=0.15\textwidth]{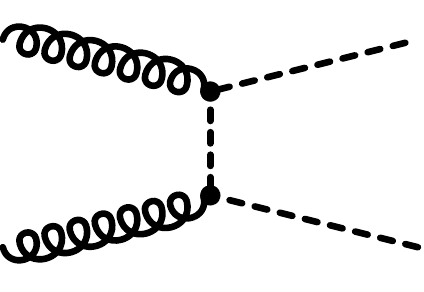}
\includegraphics[width=0.15\textwidth]{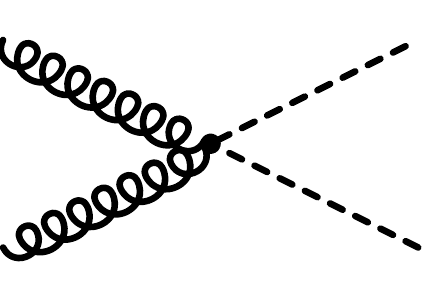}\raisebox{3ex}{$\Bigg\}$}
  \end{center}
  \caption{The diagrams contributing to squark-antisquark production in the \texttt{T2} supersymmetric simplified model.}
  \label{fig:squark_antisquark}
\end{figure}

In the simplified model of squark production \texttt{T2}, since only light-flavor squarks are present, the blob in figure~\ref{fig:t2} is represented by the diagrams for squark-antisquark production shown in figure~\ref{fig:squark_antisquark}. Squark-antisquark production occurs also through the exchange of a gluino in the  $t$-channel when a finite-mass gluino is present, as shown in figure~\ref{fig:tchannel_gluino}. With the presence of this diagram, squark pair production and mixtures of left- and right-handed squark production occur as well. That is, instead of only $\sq_\L\sq_\L^*$ production, in the case of finite $m_{\widetilde{g}}$ we also have $\sq_\L\sq_\L$, $\sq_\L\sq_\R$ and $\sq_\L\sq_\R^*$ production (as well as $\R \leftrightarrow \L$, with equal cross sections for mass-degenerate squarks).

\section{Limits for MSSM-like squarks and LSPs}
Adding a finite-mass gluino to the simplified model \texttt{T2}, we obtain what we named MSSM-like squark production. To investigate the differences in limits obtained from efficiencies of this model versus limits from efficiencies for the \texttt{T2} model in the $\smsq$-$\mne$ mass plane, we used two strongly excluding~\cite{Kraml:2013mwa,Kraml:2014sna,Beenakker:1996ch} SUSY analyses CMS-SUS-13-012~\cite{Chatrchyan:2014lfa} and CMS-SUS-12-028~\cite{Chatrchyan:2013lya}, which we named according to their main cut variables MHT and $\alpha_T$, respectively. 

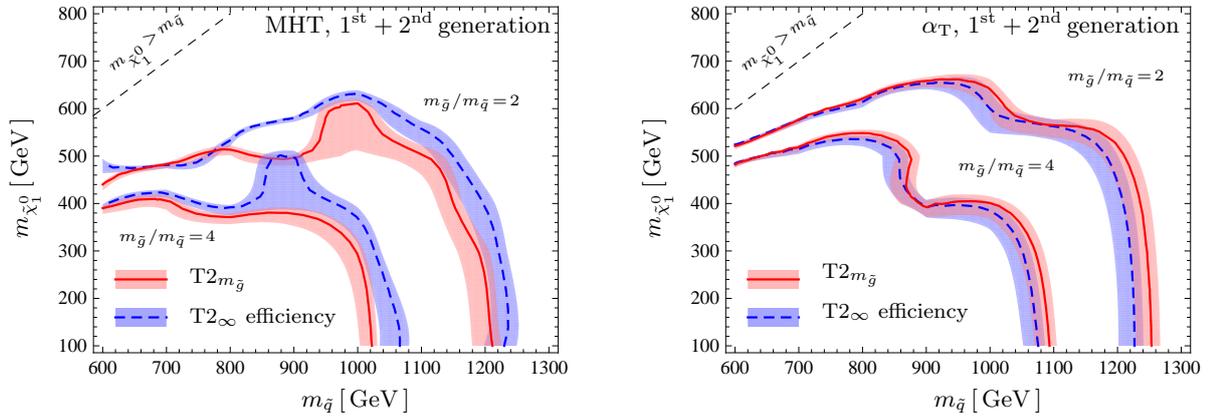
\begin{figure}[h!]
\centering
\setlength{\unitlength}{1\textwidth}
\begin{picture}(0.95,0.32)
 \put(0.52,0.008){ 
  \put(0.0,0.025){\includegraphics[scale=1.1]{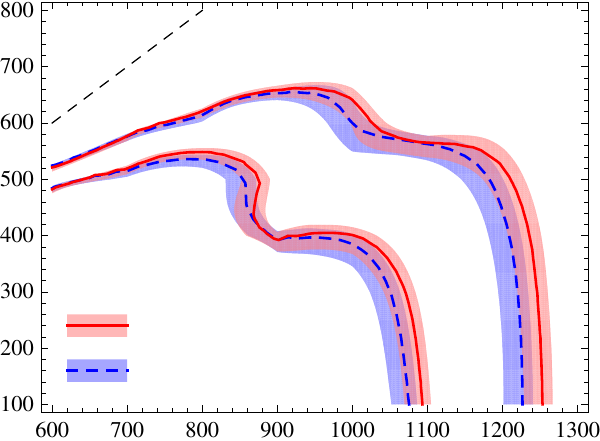}} 
  \put(0.19,0.31){\footnotesize $\alpha_{\mathrm{T}}$, $1^{\mathrm{st}}+2^{\mathrm{nd}}$\,generation}
  \put(0.2,0.0){\footnotesize $\smsq\,[\gev\,]$}
  \put(-0.04,0.14){\rotatebox{90}{\footnotesize $m_{\tilde{\chi}^0_1}\,[\gev\,]$}}
  \put(0.31,0.27){\tiny $\smgo/\smsq\!=\!2$}
  \put(0.22,0.196){\tiny $\smgo/\smsq\!=\!4$}
  %key
  \put(0.108,0.107){\scriptsize $\tmgo$}
  \put(0.108,0.074){\scriptsize $\Tinv$ efficiency}
  \put(0.04,0.278){\rotatebox{37.6}{\tiny $m_{\tilde{\chi}^0_1}\!>\!\smsq$}}
  }
\put(0,0.008){ 
  \put(0.0,0.025){\includegraphics[scale=1.1]{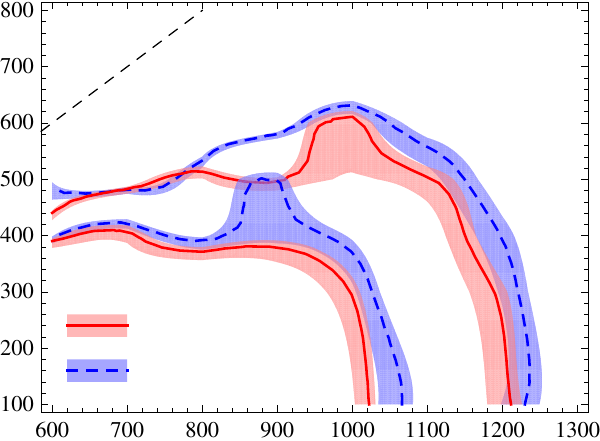}} 
  \put(0.17,0.31){\footnotesize MHT, $1^{\mathrm{st}}+2^{\mathrm{nd}}$\,generation}
  \put(0.2,0.0){\footnotesize $\smsq\,[\gev\,]$}
  \put(-0.04,0.14){\rotatebox{90}{\footnotesize $m_{\tilde{\chi}^0_1}\,[\gev\,]$}}
  \put(0.3,0.25){\tiny $\smgo/\smsq\!=\!2$}
  \put(0.05,0.137){\tiny $\smgo/\smsq\!=\!4$}
   %key
  \put(0.108,0.103){\scriptsize $\tmgo$}
  \put(0.108,0.0697){\scriptsize $\Tinv$ efficiency}
  \put(0.04,0.278){\rotatebox{37.6}{\tiny $m_{\tilde{\chi}^0_1}\!>\!\smsq$}}
  }
\end{picture}
\caption{95\% C.L. exclusion limits
  %~\cite{Edelhauser:2014ena} 
  at the 8 TeV LHC derived from the full finite-mass gluino 
$\tmgo$ 
model (red solid curves) and
from the efficiencies for the \texttt{T2}, or $\Tinv$ simplified model (blue dashed curves). The shaded regions around
the curves denote the uncertainties due to scale variation ($\mu=\smsq/2,2\smsq$). 
}
\label{fig:limssusy}
\end{figure}

Using a simulation of the 8 TeV LHC with \texttt{MadGraph 5}~\cite{Alwall:2014hca}, \texttt{Pythia 6}~\cite{Sjostrand:2006za}, and \texttt{Delphes 3}~\cite{deFavereau:2013fsa,Cacciari:2011ma} and our own implementations of the MHT and $\alpha_T$ analyses, we obtained efficiencies from which we calculated upper limits with \texttt{RooStatS}~\cite{roostatscl95}. Upon comparing these upper limits with the NLO prediction for the cross section of the MSSM-like squark production that was calculated with \texttt{Prospino}~\cite{Beenakker:1996ed}, we obtained the red solid lines in figure~\ref{fig:limssusy}~\cite{Edelhauser:2014ena}.

Repeating the process for calculating limits from efficiencies for the \texttt{T2} model, and again comparing with the NLO prediction for the cross section of the MSSM-like squark production, we obtained the blue dashed lines in figure figure~\ref{fig:limssusy}. The blue and red exclusion lines are within the error for the theoretical prediction of the cross section of the full model exclusions.

Note that the limits shown are from our own simulations and for the most sensitive bin only that was calculated with a pure background hypothesis. CMS, on the contrary, combines bins; this procedure can be expected to yield stronger exclusion limits.

\section{Production of same-spin quark partners at the LHC}
\begin{figure}[htp]\centering
\includegraphics[width=0.15\textwidth]{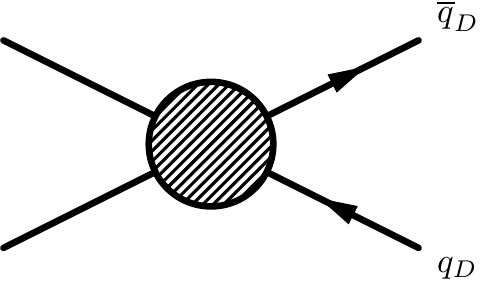}\hspace{2ex}\raisebox{3ex}{=\hspace{2ex}$\Bigg\{$}
\includegraphics[width=0.15\textwidth]{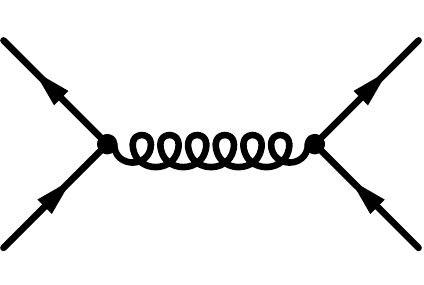}
\includegraphics[width=0.15\textwidth]{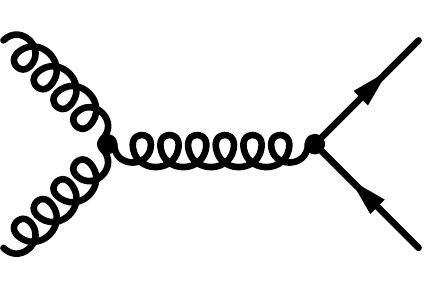}
\includegraphics[width=0.15\textwidth]{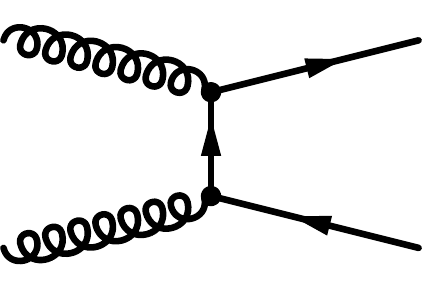}\raisebox{3ex}{$\Bigg\}$}
\caption{The diagrams contributing to KK quark pair production for the UED equivalent of the \texttt{T2} simplified model. D stands for doublet, as $q_{\mathrm{D}}$ is an SU(2) doublet quark.}
\label{fig:ued_quarks}
\end{figure}

When one assumes that in the \texttt{T2} model same-spin instead of scalar quark partners are produced, the exclusion limits from the original scalar quark partner \texttt{T2} model may change. 
In this case of same-spin KK quark production, the blob in figure~\ref{fig:t2} is for  UED-like quark production represented by the three diagrams shown in figure~\ref{fig:ued_quarks}.

%Here $q^{(1)}$ denotes a first excitation KK quark partner and D, S stand for SU(2) doublet and singlet, respectively. 
As before for MSSM-like squark production, when the gluon partner mass is finite, an additional $t$-channel diagram appears, yielding not only KK quark-antiquark production 
$q_{\mathrm{D}}\bar{q}_{\mathrm{D}}$ %and $q_{\mathrm{S}}\bar{q}_{\mathrm{S}}$
(where D, S stand for SU(2) doublet and singlet, respectively),
but also KK quark pair and mixed doublet and singlet production $q_{\mathrm{D}}q_{\mathrm{D}}$, $q_{\mathrm{D}}q_{\mathrm{S}}$ and $q_{\mathrm{D}}\bar{q}_{\mathrm{S}}$ 
(and $\mathrm{D}\leftrightarrow\mathrm{S}$ with equal cross sections).
%$q_i^{(1)}\bar{q}_i^{(1)}$, $i\in\{\mathrm{D},\mathrm{S}\}$ but also production of $q_i^{(1)}{q}_j^{(1)}$ and $q_i^{(1)}\bar{q}_j^{(1)}$, with $i,j\in\{\mathrm{D},\mathrm{S}\}$.
%$q_{\mathrm{D}}q_{\mathrm{D}}$, $q_{\mathrm{D}}q_{\mathrm{S}}$ and $q_{\mathrm{D}}\bar{q}_{\mathrm{S}}$.

\section{Limits for UED-like quarks and LKPs}
\begin{figure}[]
  \begin{center}
    \includegraphics[width=0.47\textwidth]{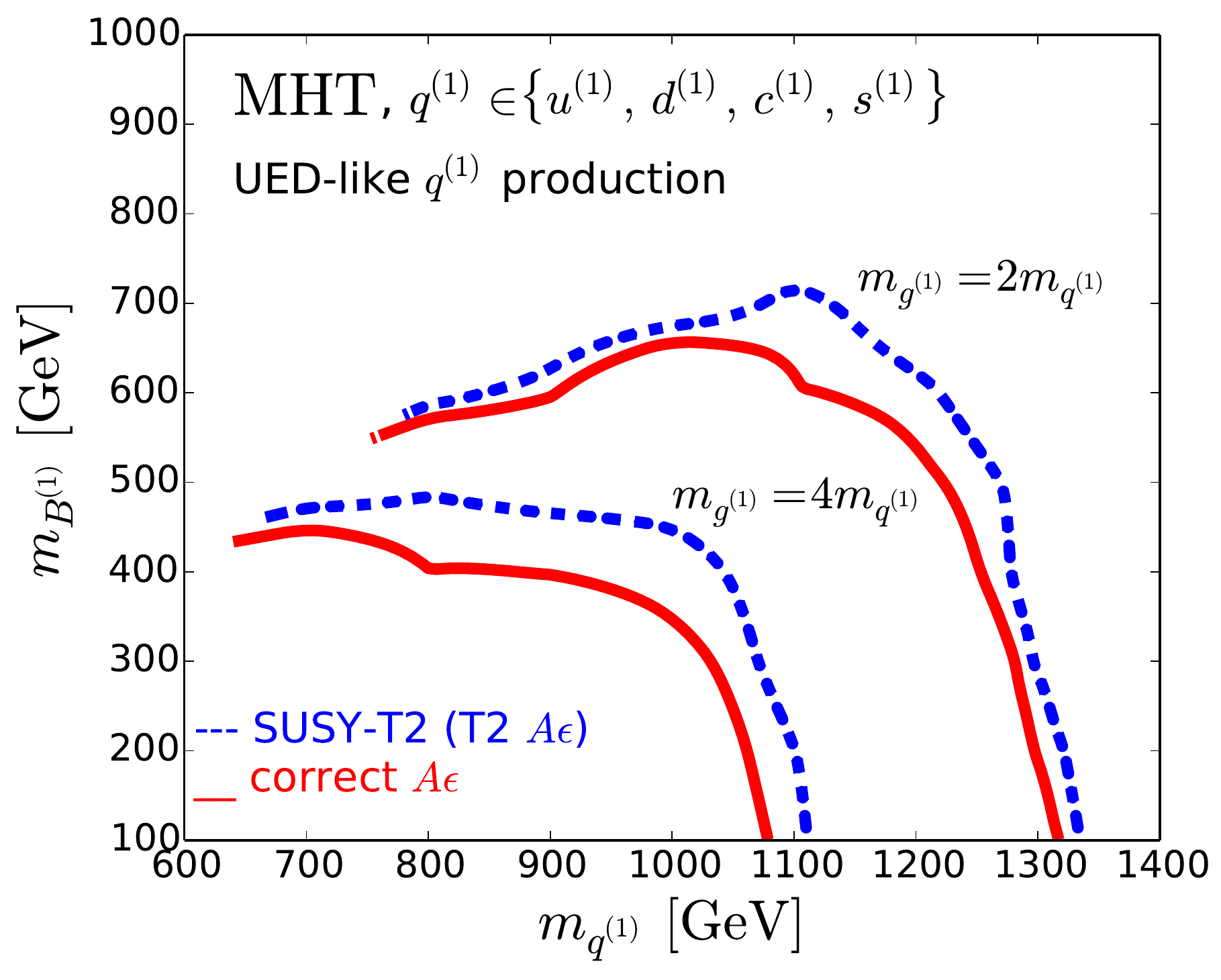}
    \includegraphics[width=0.47\textwidth]{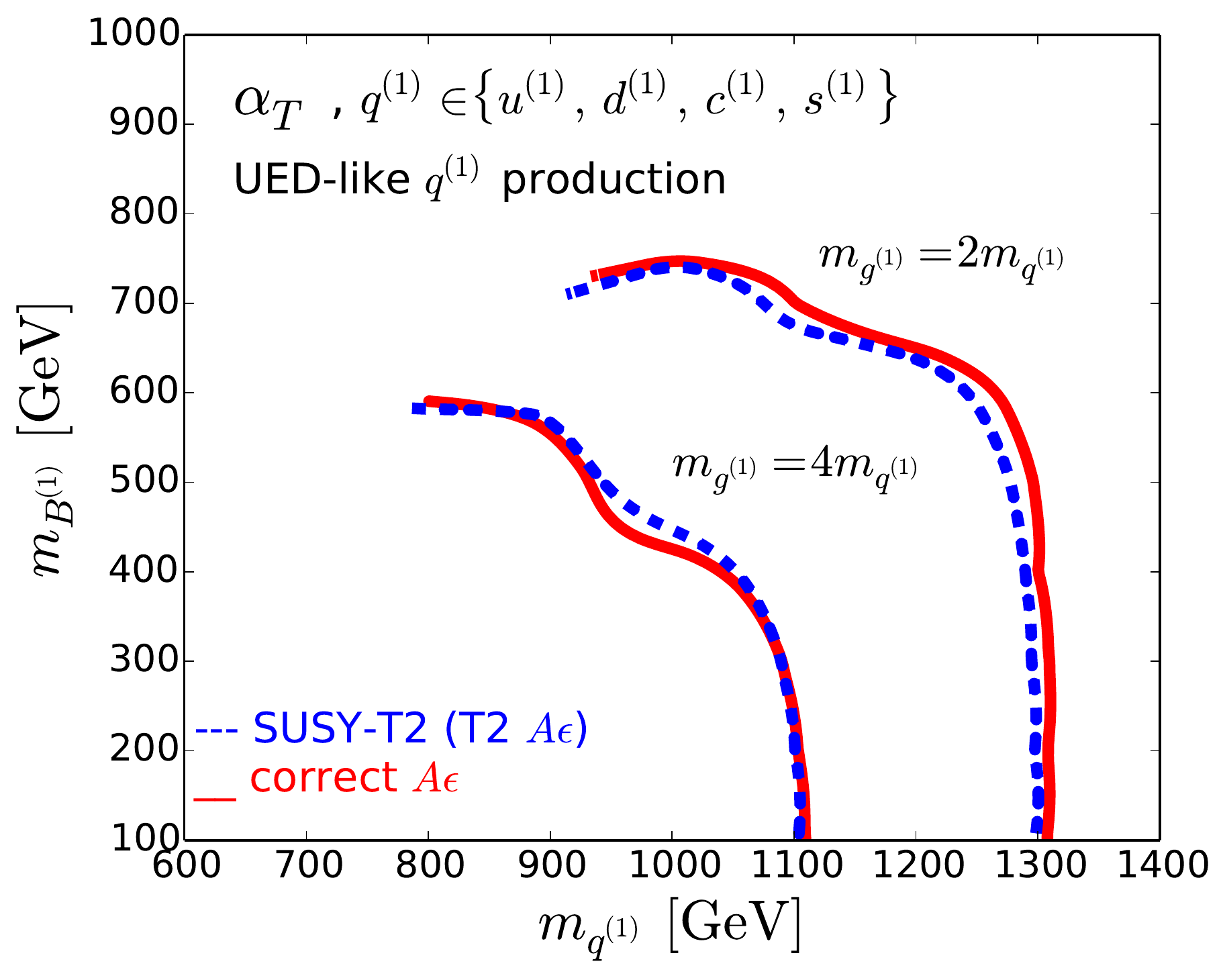}
  \end{center}
  \caption{Limits for a model of UED-like quark production at the 8 TeV LHC from the simplified SUSY-\texttt{T2} model (blue dashed line) and from a simulation of the full model (red solid line). 
  Same-spin quark partners are denoted $q^{(1)}$; the lightest KK particle is denoted $B^{(1)}$.
  }
  \label{fig:limsued}
\end{figure}

We investigated the difference in limits obtained for a model of UED-like quark production from the full model, containing spin-1/2 quark partners and a finite gluon partner mass, as opposed to limits for this model from the SUSY-\texttt{T2} model containing scalar quark partners and no gluon partner. Following the same procedure as described in the previous section, with now \texttt{MadGraph 5} \@LO cross section predictions of UED-like quark production, we again find that the limit curves in the quark partner and LKP mass plane are close~\cite{Edelhauser:2015ksa}, as shown in figure~\ref{fig:limsued}. The $\alpha_T$ analysis slightly underestimates, whereas the MHT analysis overestimates the limits. 

Note that for this model, the limits are up to 1300 GeV quark partner masses. Most experimental results up to now, however, present limits up to 1 TeV squark masses. This means that when a BSM model with higher cross sections than a simplified SUSY model is tested against those results, a tool like \texttt{SModelS} would not contain and hence not yield any results in these higher mass regions.

\section{Conclusion}
We conclude that simplified models are a good approximation for (1) more general models of supersymmetry and (2) same-spin models that are not originally described by these models~\cite{Edelhauser:2014ena,Edelhauser:2015ksa}.

Since recasting simplified model results is faster than a complete analysis of a specific model or calculating efficiencies for that model, they are ideal for performing global fits using e.g. a combination of \texttt{SModelS}~\cite{Kraml:2013mwa,Kraml:2014sna} and~\texttt{Fittino}~\cite{Bechtle:2013mda}.

\section*{Acknowledgments}

I thank the organizers of the 50th Rencontres de Moriond for giving me the opportunity to present these results.
I thank Lisa Edelh\"{a}user, Michael Kr\"{a}mer, Jan Heisig, and Lennart Oymanns for collaboration on this work.
I also thank Christian Autermann and Wolfgang Waltenberger for useful discussions and suggestions, as well as the Institute of High Energy Physics (HEPHY) for their hospitality.
The work described here was supported by the Deutsche 
Forschungsgemeinschaft through the graduate school ``Particle and Astroparticle Physics in the Light of the LHC''.

\end{document}